\documentclass{JINST}
\let\ifpdf\relax

\usepackage{float}
\usepackage{amsmath, amsthm, amssymb}

\title{A study of electron recombination using highly ionizing particles in the ArgoNeuT Liquid Argon TPC}

\author{R. Acciarri$^a$, C. Adams$^b$, J. Asaadi$^c$, B. Baller$^a$$^,$\thanks{Corresponding
author.}, T. Bolton$^d$, C. Bromberg$^e$, \linebreak
F. Cavanna$^b$$^,$$^f$, E. Church$^b$, D. Edmunds$^e$, A. Ereditato$^g$, S. Farooq$^d$, B. Fleming$^b$, \linebreak
 H. Greenlee$^a$, G. Horton-Smith$^d$, C. James$^a$, E. Klein$^b$, K. Lang$^h$, P. Laurens$^e$, \linebreak
 D. McKee$^d$, R. Mehdiyev$^h$, B. Page$^e$, O. Palamara$^b$$^,$$^i$, K. Partyka$^b$, G. Rameika$^a$, \linebreak
B. Rebel$^a$, M. Soderberg$^a$$^,$$^c$, J. Spitz$^b$, A.M. Szelc$^b$, M. Weber$^g$, M. Wojcik$^j$, T. Yang$^a$, G.P. Zeller$^a$\\
\llap{$^a$}Fermi National Accelerator Laboratory,  Batavia, IL  60510 USA\\
\llap{$^b$}Yale University,  New Haven, CT  06520  USA\\
\llap{$^c$}Syracuse University,  Syracuse, NY  13244  USA\\
\llap{$^d$}Kansas State University,  Manhattan, KS  66506 USA\\
\llap{$^e$}Michigan State University, East Lansing, MI  48824 USA\\
\llap{$^f$}Universit\'a dell'Aquila e INFN,  L'Aquila, Italy\\
\llap{$^g$}University of Bern,  Bern, Switzerland\\
\llap{$^h$}The University of Texas at Austin,  Austin, TX  78712  USA\\
\llap{$^i$}INFN - Laboratori Nazionali del Gran Sasso,  Assergi, Italy\\
\llap{$^j$}Lodz University of Technology,  Lodz, Poland\\

 E-mail:\email{baller@fnal.gov}}

\abstract{Electron recombination in highly ionizing stopping protons and deuterons is studied in the ArgoNeuT detector. The data are well modeled by either a Birks model or a modified form of the Box model. The dependence of recombination on the track angle with respect to the electric field direction is much weaker than the predictions of the Jaffe columnar theory and by theoretical-computational simulations.}


\begin{document}

\section{Introduction}

Liquid Argon Time Projection Chambers (LAr TPCs) offer excellent calorimetry and mm-scale position resolution in a large volume and are increasingly favored for the next generation of neutrino detectors. Particles in the MeV - few GeV energy range of interest for neutrino detectors are usually contained within the TPC and can be well characterized by range, calorimetry and decay topology. Calorimetric reconstruction requires calibration to correct for detector specific effects such as liquid argon impurities and electronics response and calibration for charge loss due to electron-ion recombination.

In this paper, we report on a study of the recombination of electron-ion pairs produced by ionizing tracks in a sample of stopping protons and deuterons in the ArgoNeuT detector with a particular emphasis on the angular dependence. The data were taken during an exposure of 1.35x10$^{20}$ protons on target in the Fermilab NuMI neutrino beam\cite{numi}.

\section{Recombination Models}
\label{models}

Electrons emitted by ionization are thermalized by interactions with the surrounding medium after which time they may recombine with nearby ions. The Onsager geminate theory\cite{onsager} presumes that the dominant recombination process is the re-attachment of the electron to the parent ion under the influence of the Coulomb field of the pair. In the columnar model of Jaffe\cite{jaffe}, published in 1913, recombination depends on the collective electron and ion charge density from multiple ionization interactions in a cylindrical volume surrounding the particle trajectory. 

The relative importance of these two theories for liquid argon can be estimated by comparing the average electron-ion distance with the average ion-ion distance. The average separation distance between ions, $r_{ion}$, is $W_{ion}/(dE/dx)$ where $W_{ion}$ = 23.6 eV is the energy required to ionize an argon atom. 
For the range of particle stopping power, $dE/dx$, in this analysis, $r_{ion}$ varies between 10 and 50 nm. Transient conductivity measurements of ionization produced by charged particles show that electrons reach thermal energies in 1 - 2 ns \cite{sowada} in liquid argon during which time the electrons are estimated to travel $\mathcal{O}(10^3)$ nm. As a result, the probability of geminate recombination should be small since the electron-ion separation distance is much larger than the ion-ion separation distance. A simulation of thermalization and recombination in \cite{wojcik} estimates that the probability of geminate recombination is  $\sim10^{-3}$. These studies favor a columnar theory approach and imply that collective effects are important. 

In the Jaffe theory, electrons and ions are assumed to have a Gaussian spatial distribution around the particle trajectory during the entire recombination phase. The spatial distribution and the charge mobility, $\mu$, are assumed to be equal for electrons and ions. With these assumptions, Jaffe found that the fraction that survive recombination is

\begin{equation}
\label{jaffe}
{\mathcal R_{\rm J}} = \left[ {1 + \frac{\alpha N_o}{8 \pi D} \sqrt{\frac{\pi}{z'}} S(z') } \right] ^{-1} \quad \mathrm{where}
\quad z' = \frac{b^2 \mu^2 \mathcal{E}^2 sin^2 \phi }{2 D^2},
\end{equation}

\noindent
where $\alpha$ is a recombination coefficient, $N_o$ is the number of ion-electron pairs per unit length, $D$ is the diffusion coefficient, $\mathcal{E}$ is the electric field and $b = r_o \sqrt{4/\pi}$ where $r_o$ is the average ion-electron separation distance after thermalization. The variable $\phi$ is the angle between the electric field and the particle direction. Equation \ref{jaffe} is now referred to as Birks law \cite{birk} which was developed to model the scintillation light yield of a particle as a function of the stopping power. The angular dependence inherent in the dimensionless variable $z'$ may be modified by the $S(z')$ factor,

\begin{equation}
\label{szprime}
S(z') = \frac{1}{\sqrt\pi{}} \int_0^\infty \frac{e^{-s} \, ds}{\sqrt{s (1 + s /z')} }.
\end{equation}

\noindent
In this equation, $s$ is a dimensionless variable that characterizes the time dependent overlap of the electron and ion distributions. Numerical integration of equation \ref{szprime} shows that $S(z')$ approaches one when $z'$ exceeds 10.  To estimate $z'$ for this analysis, we assume that electrons are thermalized and use the Einstein-Smoulchowski relation $D / \mu \approx E_{thermal} \approx$ 0.01 eV for liquid argon. We set $\mathcal{E}$ to a low electric field typically used in a LAr TPC (0.5 kV/cm), set $r_o$ = 2500 nm\cite{wojcik} and find that $z'$ is typically > 40 for the range of angles in this analysis. The variation in $S(z')$ is less than 1\% in this range, allowing the use of a more compact form

\begin{equation}
\label{columnar}
{\mathcal R_{\rm J}} = \left[ {1 + k_c (dE/dx) / (\mathcal{E} sin \phi) } \right]^{-1},
\end{equation}

\noindent
where $k_c$ is a constant that is assumed to be specific to liquid argon. 

Thomas and Imel \cite{thomas} noted that electron diffusion and ion mobility are negligible in liquid argon during recombination. After dropping these terms in the Jaffe diffusion equations and applying ``Box model'' boundary conditions they find

\begin{equation}
\label{box}
{\mathcal R_{\rm{Box}}} = \frac{1}{\xi}ln(\alpha + \xi), \quad  \mathrm{where} \quad \xi = k_{\rm{Box}} N_o / 4a^2\mu\mathcal E.
\end{equation}

\noindent
The quantity $N_o/4a^2$ represents the charge density in a microscopic box of size $a$. The variable $\alpha$ is explicitly equal to one in the canonical form of the model. The model was developed to parameterize the electric field dependence of recombination using radioactive sources where the charge density in the box is a fixed quantity. 
The similarity between the Box model and the Jaffe theory becomes clear by performing a series expansion of equation \ref{box}, $ln(1+\xi)/\xi = 1/(1 + \xi/2 ...)$. The Jaffe and Box recombination $k$ factors are related by a  multiplicative factor when $\xi$ < 1. 

The electric field is held constant in this analysis but the charge density varies along a track. We therefore associate the charge density with $dE/dx$ and re-cast $\xi$ in the  form

\begin{equation}
\label{xidef}
\xi = \beta (dE/dx),
\end{equation}

\noindent
where $\beta$ is a parameter found by performing a recombination fit.  There is no explicit mention of angular dependence in the Box model. It is generally accepted that $\mathcal{E}$ can be replaced by $\mathcal{E} sin \phi$.

\subsection{Theoretical application}
\label{sec:application}

Ideally, a theoretical understanding of recombination would result in a prescription for calorimetric reconstruction of particles in a liquid argon TPC. The Birks and Box model equations do not provide a consistent global description of all data, which is not remarkable considering the assumptions made in their development.  They do however provide good agreement with data in some regimes, for example radioactive source data where $dE/dx$ is fixed and the electric field is varied. ICARUS\cite{icarrecomb} was the first experiment to study recombination with variable $dE/dx$ and variable $\mathcal{E}$ in the range of interest for neutrino liquid argon TPCs. The ICARUS data are well described by a Birks form similar to equation \ref{columnar}:

\begin{equation}
\label{birks}
\mathcal{R}_{\rm{ICARUS}}  = \frac{A_B}{1 + k_B\cdot (dE/dx)/\mathcal{E}}
\end{equation}

\noindent
with $A_B$ = 0.800 $\pm$ 0.003 and $k_B$ = 0.0486 $\pm$ 0.0006 (kV/cm) (g/cm$^2$)/MeV. Note that in the high electric field limit, $\mathcal{R}_{\rm{ICARUS}} \rightarrow 0.8$ whereas the canonical box model would predict $\mathcal{R}_{\rm{Box}} \rightarrow 1$. The recombination factor approaches 0 for the canonical forms of both models. This expected behavior at small electric field is supported by measurements with relativistic heavy ions ($\mathcal{R} \sim 0.003$) but not with low energy electrons ($\mathcal{R} \sim 0.35$)\cite{doke}. 

A technical difficulty arises when applying a Birks model correction to highly ionizing particles. Calorimetric reconstruction of particle tracks relies on the equation

\begin{equation}
\label{dedxeq}
dE/dx =  (dQ/dx) / (\mathcal{R} \; W_{ion}),
\end{equation}

\noindent
to determine the stopping power given a measured value of the charge deposited per unit length, $dQ/dx$, along the particle trajectory. The result, using the Birks model form for $\mathcal{R}$ is 

\begin{equation}
\label{birksinv}
dE/dx = \frac{dQ/dx}{A_B / W_{ion} - k_B \cdot (dQ/dx)/\mathcal{E}}.
\end{equation}

\noindent
Spurious values of $dE/dx$ result when the denominator approaches zero (at large values of $dQ/dx$). 
In contrast, the inverse Box equation has an exponential form that does not suffer from this malady: 

\begin{equation}
\label{boxinv}
dE/dx = (exp(\beta W_{ion} \cdot (dQ/dx)) - \alpha)/\beta.
\end{equation}

In summary, the Birks model provides a consistent description of recombination over a limited range of $dE/dx$ but there are technical difficulties applying it to calorimetric reconstruction at high ionization. The Box model has no technical difficulties but has inadequate behavior at low $dE/dx$. 

\begin{figure} [H]
\centering
\includegraphics [width = 0.5\textwidth] {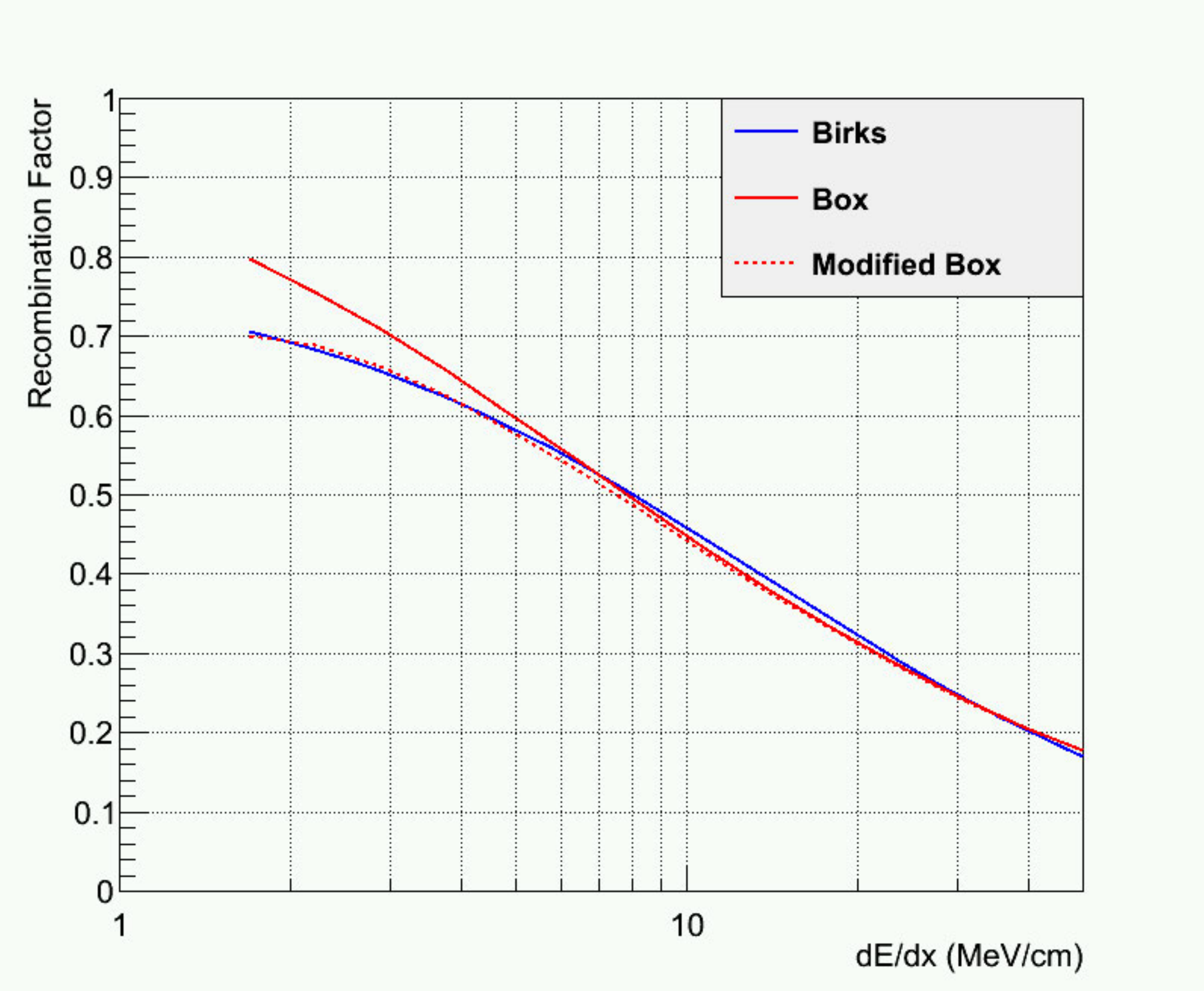}
\caption{Qualitative comparison of the recombination factor for the Birks equation using ICARUS parameters (blue curve), the ``canonical'' Box model with $\alpha$ = 1 and $\beta$ = 0.30 (red curve) and a ``modified Box" equation with $\alpha$ = 0.93 and $\beta$ = 0.30 cm/MeV (dotted red curve). }
\label{fig:birkbox}
\end{figure}

This deficiency in the Box model can be corrected by allowing $\alpha$ < 1 as illustrated in figure \ref{fig:birkbox}. The recombination factor using the Birks model with ICARUS parameters is shown by the blue curve with an electric field of 0.5 kV/cm and liquid argon density of 1.383 g/cm$^3$. One can adjust the canonical Box model $\beta$ value to 0.30 cm/MeV to match the blue curve at $dE/dx$ = 7 MeV/cm  and achieve good qualitative agreement to very high values of stopping power (solid red curve). The significant disagreement at low $dE/dx$ can be eliminated by setting $\alpha$ = 0.93 while keeping $\beta$ = 0.30 cm/MeV (dotted red curve). In this report we will refer to this case as a ``modified Box'' model.

In the analysis of ArgoNeuT data, we perform fits to the Birks model recombination parameters $A_{\rm{Argo}}$ and $k_{\rm{Argo}}$ in different angular bins, and to the modified Box model recombination parameters $\alpha$ and $\beta$ in different angular bins to test the validity of the $\phi$ dependence predicted by the columnar theory.

\section{Calorimetric Reconstruction in the ArgoNeuT Detector}
\label{sec:caloreco}

The ArgoNeuT detector is described in reference \cite{argoneut}. All data were taken with an electric field of 0.481 kV/cm. The cryostat was operated slightly above atmospheric pressure, keeping the liquid argon density at 1.383~g/cm$^3$. 

The calorimetric reconstruction of 3D tracks is described in reference \cite{muonpaper}. Proximal hits in each wire plane are grouped together into line-like two dimensional clusters. Clusters in each wire plane, or view, are matched to form three-dimensional (3D) tracks comprised of a set of space points. Each space point consists of a 3D position in the detector. Calorimetric measurements are made using the charge collected by the collection plane. A correction is made for charge loss due to attachment on impurities in the liquid argon while the electrons drift from the ionizing track to the wire plane.  The result is a measurement of the charge $dQ$ remaining at the space point position after recombination. $dQ/dx$ is  found by projecting the track into the coordinate system of the collection plane. A calorimetric measurement of the stopping power, $(dE/dx)_{calo}$, is then found for each space point by applying a recombination correction using equation \ref{boxinv}.

On average, a track stops half-way through the last wire cell, i.e. directly above the last collection plane wire, leading to an uncertainty in the stopping point position equal to the space point separation / $\sqrt{12}$. The space point separation varies between 0.4 - 1 cm for tracks in this analysis, resulting in an average uncertainty of 0.2 cm. Some improvement in estimating the position of the stopping point can be gained by inferring the stopping point within the last cell, denoted $\Delta$, from the pattern of energy loss using the last four space points. $\Delta$ is varied in 0.1 cm increments (0.1 cm < $\Delta$ < 1 cm). $(dE/\Delta)_{calo}$ of the last point and $(dE/dx)_{hyp}$ for the last four space points are calculated using this value of $\Delta$. $(dE/dx)_{hyp}$ is defined in the next section. The rms difference between $(dE/dx)_{calo}$ and $(dE/dx)_{hyp}$ is found for each value of $\Delta$.  The value of $\Delta$ with the smallest rms value is chosen as the stopping point position. The estimated stopping point error from the Monte Carlo using this procedure is 0.1 cm.
The residual range of all space points are corrected for the $\Delta$ offset.

\section{Stopping Particle Identification}

Unlike the situation in a test beam where the incident particle type and energy is known, a wide variety of particles are produced by the neutrino beam in a wide range of energies. The best discriminant for stopping particle identification uses calorimetry, which  requires making a recombination correction - the focus of this study. In contrast to previous studies that rely on a Monte Carlo simulation, we utilize the theoretical power-law dependence of $dE/dx$ on the particle velocity as it reaches the end of its travel. We introduce a particle identification technique using this feature that is intended to minimize potential sources of selection bias.

\begin{figure} [h!]
\centering
\includegraphics [width = 0.6\textwidth] {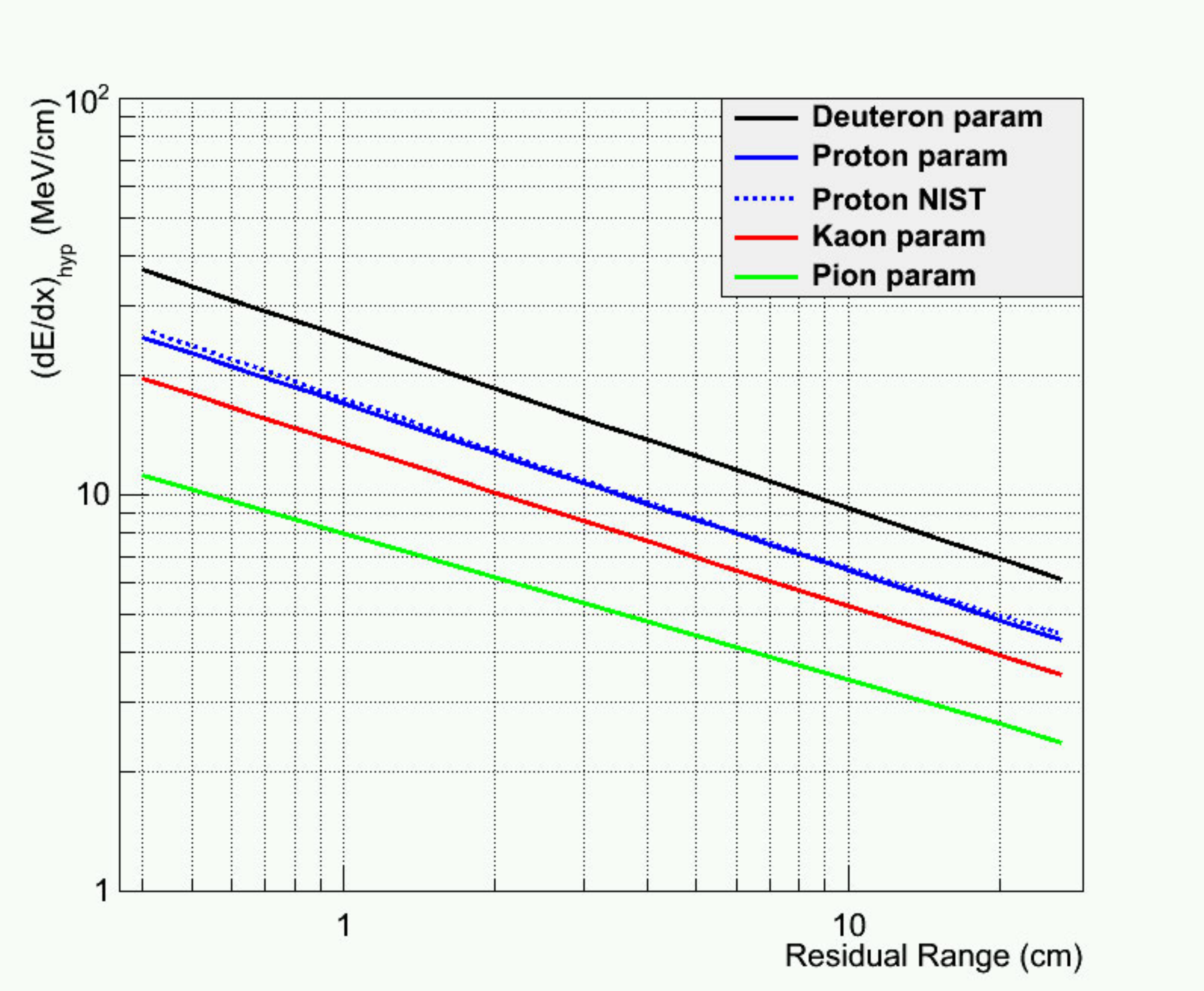}
\caption{Parameterized stopping power for various particles as a function of residual range (solid lines). The blue dotted curve shows the NIST stopping power for protons in argon at a density of 1.383 g/cm$^3$.}
\label{fig:bethebloch}
\end{figure}

The method relies on numerically tracking a particle of mass $M$ with momentum $p$ in steps of length $dx$. The initial kinetic energy, $T_0$, of the particle is specified with the condition $p/M$ < 1 to ensure that it is in the range of applicability for the power-law dependence.  The energy lost by ionization in the first step is found by multiplying the stopping power as calculated from the Bethe-Bloch equation\cite{bethebloch} by the step length. The shell correction and density correction terms are small in this range and are neglected. The kinetic energy of the second step is set to $T_1 = T_0 - (dE/dx) dx$. This process is repeated until the energy lost in the last step exceeds the kinetic energy of the previous step. This procedure is equivalent to a tracking algorithm in a Monte Carlo simulation without the complicating contributions from energy loss fluctuations and $\delta$-rays. The stopping power is then parameterized as a function of the ``residual range'', R, defined as the distance between a point on the idealized track and the stopping point. The results, shown in figure \ref{fig:bethebloch},  display the power law relationship of Equation \ref{dedxpar} for a variety of particle masses. The values $A$ = 17 and $b$ = -0.42 provide a good representation of the NIST projected range table\cite{nist} for protons with an uncertainty less than 3\% for kinetic energy between 50 MeV and 300 MeV (blue dotted line). In this report $(dE/dx)_{hyp}$ denotes the idealized stopping power behavior expected for a particle type hypothesis. Integration of $(dE/dx)_{hyp}$ over the extent of the particle travel yields a parameterization for the kinetic energy, $T_{range}$, equation \ref{eq:Trange}.  Table \ref{paramtable} shows the power law parameterization for particles of interest in this analysis.

\begin{eqnarray}
\label{dedxpar}
(dE/dx)_{hyp} & = & A~R^{b}  \\
T_{range} & = & \frac{A}{b+1}~R^{b+1} \label{eq:Trange}
\end{eqnarray}

\begin{table} [h!]
\centering
\begin{tabular} {| c | c | c | }
\hline
Particle    &  $A$   & $b$  \\
              &    MeV/cm$^{1-b}$ & \\                            
\hline \hline
pion        &  8  &  -0.37   \\
kaon       & 14 &  -0.41   \\
proton    & 17  & -0.42   \\
deuteron & 25   & -0.43    \\
\hline 
\end{tabular}
\caption{Stopping power parameterization for various particle types in argon at a density of 1.38 g/cm$^3$.}
\label{paramtable}
\end{table}

\begin{figure} [h!]
\centering
\includegraphics [width = 0.6\textwidth] {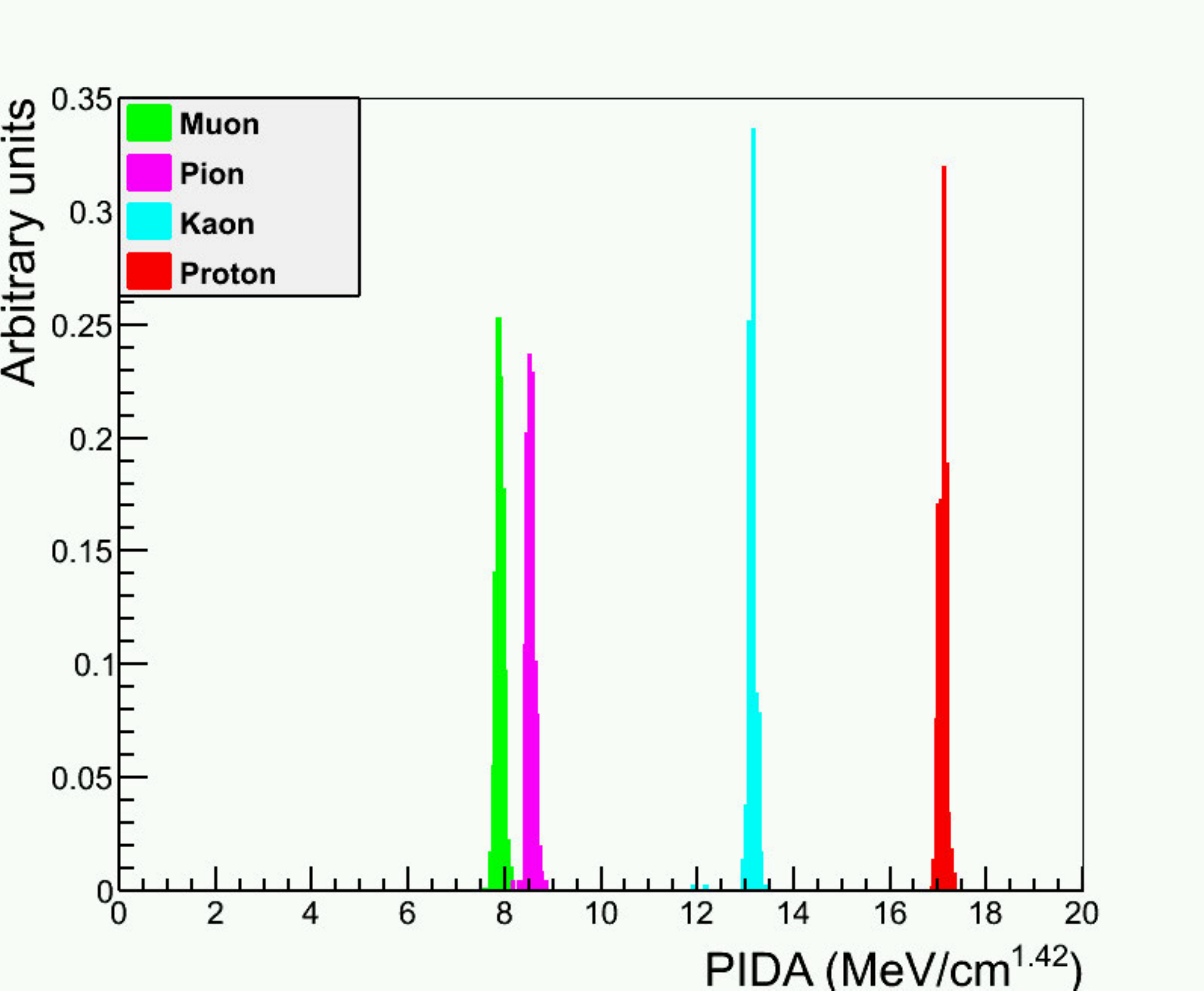}
\caption{PIDA distribution for Monte Carlo generated muons, pions, kaons and protons using truth information.}
\label{fig:pida_true}
\end{figure}

The observation that all particle types have a weak dependence on $b$ motivates constructing a particle identification algorithm based on equation \ref{dedxpar}, referred to here as PIDA. We begin by setting $b$ = -0.42. With the measurement of $(dE/dx)_{calo}$ and the residual range for each space point $i$, an estimate for A is obtained for each space point: $A_i = (dE/dx)_{calo,i} R_i^{0.42}$. The particle identification variable PIDA is defined to be the average of $A_i$ over all space points on the track. Particles may then be identified by making a histogram of PIDA for all tracks and comparing the peak values with the values of $A$ in table \ref{paramtable}. The error on PIDA introduced by fixing the value of $b$ is small compared to the error from ionization fluctuations. The method is illustrated in figure \ref{fig:pida_true} for muons, pions, kaons and protons simulated with GEANT4\cite{geant}.  In this figure, PIDA was calculated using the truth values of the energy deposition, tracking step size and residual range. Significant broadening of these distributions will occur after detector resolution and reconstruction effects are included.

In contrast to a $\chi^2$ test to a particle hypothesis, this method does not require a Monte Carlo simulation that is well-matched to the data to estimate the error used in the $\chi^2$ test. Instead, we fit a histogram of PIDA to one or more Gaussian distributions and use the fit values to guide the particle selection region. If, for example, there is an overall scale error in the detector calibration, the mean values of the Gaussian peaks will be shifted from the expected values of A. Likewise, use of an incorrect recombination correction will shift the peaks and possibly broaden the PIDA distributions.

\section{Data Selection}

There are several sources of protons that will stop in the detector. Fully contained protons produced by neutrino interactions in the detector are an obvious source. Care is taken to ensure that track mis-reconstruction near the neutrino interaction vertex does not affect calorimetric reconstruction. Neutrino or muon interactions in material surrounding the detector also produce charged particles that can enter the detector and stop. These entering tracks are more isolated but care must be taken to ensure that the detector response is uniform within the fiducial volume. We have found that production by a third mechanism is dominant - neutron interactions on argon nuclei.

\begin{figure} [h]
\centering
\includegraphics [width = 0.47\textwidth] {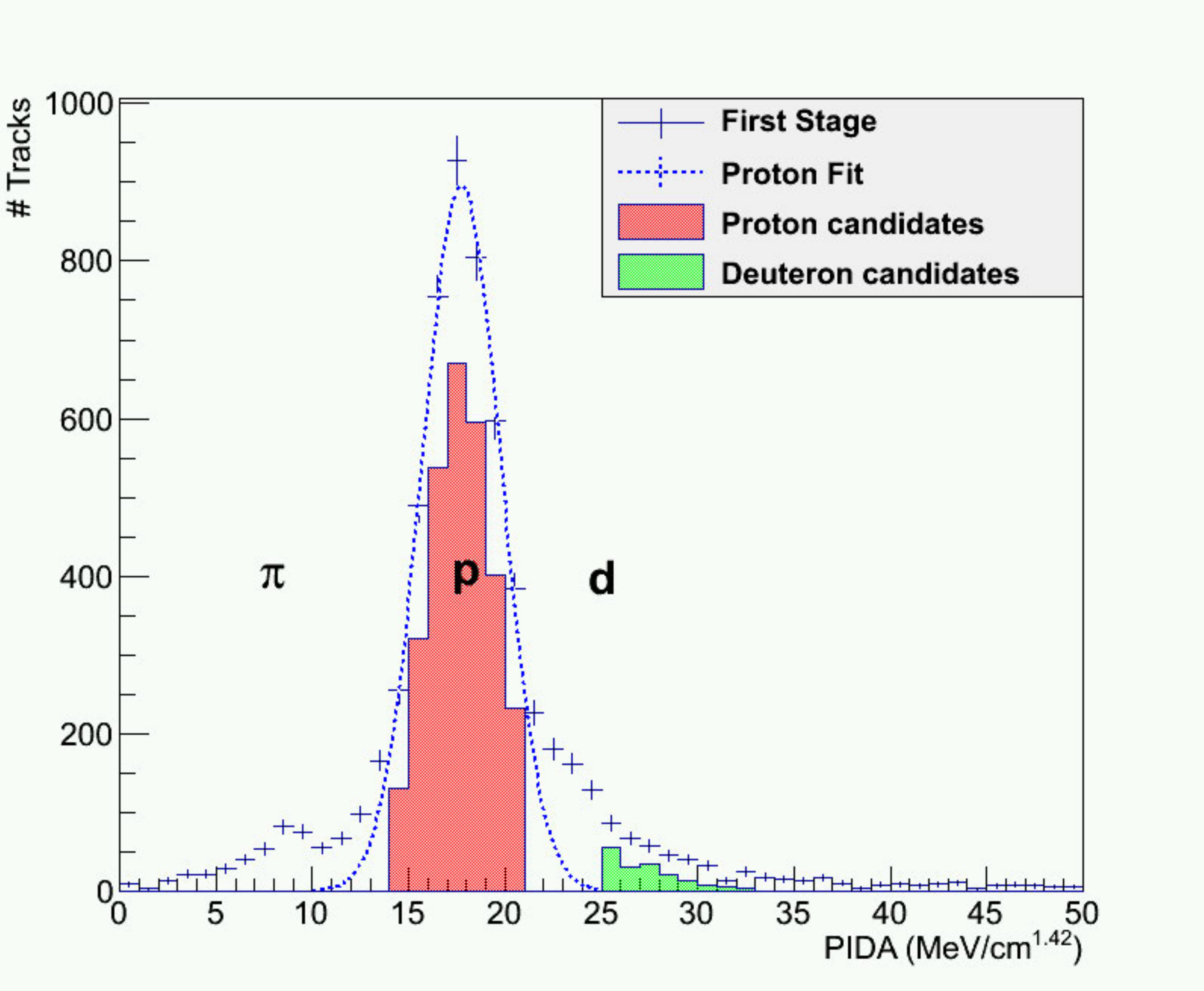}
\includegraphics [width = 0.47\textwidth] {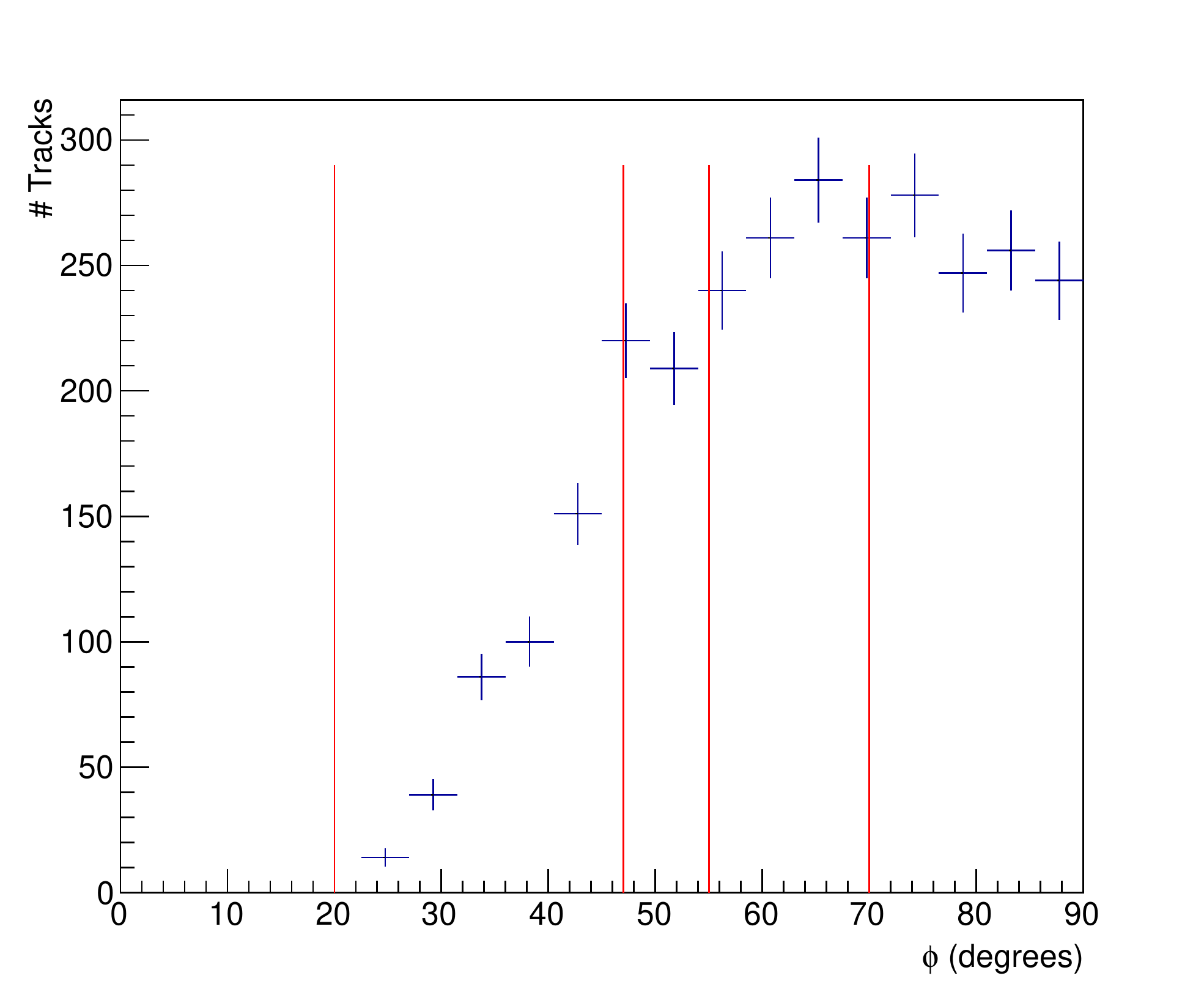}
\caption{Left: PIDA distribution for tracks passing the cuts described in the text. The expected values of PIDA for pions, protons and deuterons are indicated by the text. The Gaussian curve is a fit to the data points within the proton selection region. The shaded histograms show the sample of proton and deuteron candidates that survive all cuts and visual inspection. Right: The $\phi$ angular distribution for protons. The boundaries of the 40$^\circ$, 50$^\circ$, 60$^\circ$ and 80$^\circ$ angle bins are shown in red.}
\label{fig:pida}
\end{figure}

Data analysis and simulations are done within the framework of the LArSoft package. A stopping track is identified by requiring that the end with the higher average charge deposition, the stopping point, is more than 5 cm away from the boundary of the detector. Tracks entering the detector are not rejected. Tracks are required to have at least 6 contiguous and unique collection plane hits; i.e. tracks are rejected if collection plane hits are used in more than one space point or if a hit is not reconstructed on every collection plane wire between the two ends of the track. 
Candidate stopping protons are identified by requiring that the total energy deposited from calorimetric reconstruction, $T_{calo}$, have at least 70\% of the energy expected from the particle range, $T_{range}$, assuming the particle is a proton. The PIDA distribution of tracks passing the first stage cuts is shown in figure \ref{fig:pida} (data points with statistical errors). 

A large peak is observed at PIDA = 18 which we identify as proton tracks. A smaller bump at PIDA $\approx$ 8 is consistent with a muon or pion hypothesis. We attribute the excess of events at higher PIDA to the presence of deuterons.

In the second stage of analysis, proton candidates are selected by requiring 14 < PIDA < 21 MeV/cm$^{1.42}$. A deuteron candidate selection region, 25 < PIDA < 33 MeV/cm$^{1.42}$, is offset from the expected value of PIDA to eliminate proton contamination. All tracks in the selection regions are visually scanned. Tracks in these PIDA ranges are considered proton or deuteron candidates if all hits are perfectly reconstructed.
The resulting sample of tracks after applying all cuts is 2900 proton candidates and 170 deuteron candidates. The shaded histograms in figure \ref{fig:pida} show the PIDA distributions of these samples. The $\phi$ distribution of the proton candidates is also shown in figure \ref{fig:pida}. Tracks oriented along the neutrino beam direction populate the region $\phi \approx$ 90$^\circ$. The number of tracks at small $\phi$ is reduced by two effects; a smaller production rate and a lower reconstruction efficiency in the induction plane. The latter effect is an innate feature of the liquid argon TPC.

In general, the tracks are isolated and are not associated with a neutrino or muon interaction in the detector volume. The number of protons is  $\sim$30 times higher than that expected from neutral current neutrino interactions. The dominant production mechanism is likely the $^{40}$Ar(n,p)$^{40}$Cl reaction from neutrons produced by neutrinos and muons in the MINER$\nu$A detector that is located upstream of ArgoNeuT. A Fluka\cite{fluka} simulation of 0.5 - 1 GeV neutrons interacting in liquid argon supports this conjecture. The simulation predicts a large production rate for single protons. The simulation also predicts that deuteron production, $^{40}$Ar(n,d)$^{39}$Cl, is $\sim$1/4 of the proton production rate, while triton production is a factor of 10 less than the deuteron rate. These are rough estimates since the neutron energy spectrum of the NuMI beam is not known.

\begin{figure} [h!]
\centering
\includegraphics [width = 0.8\textwidth] {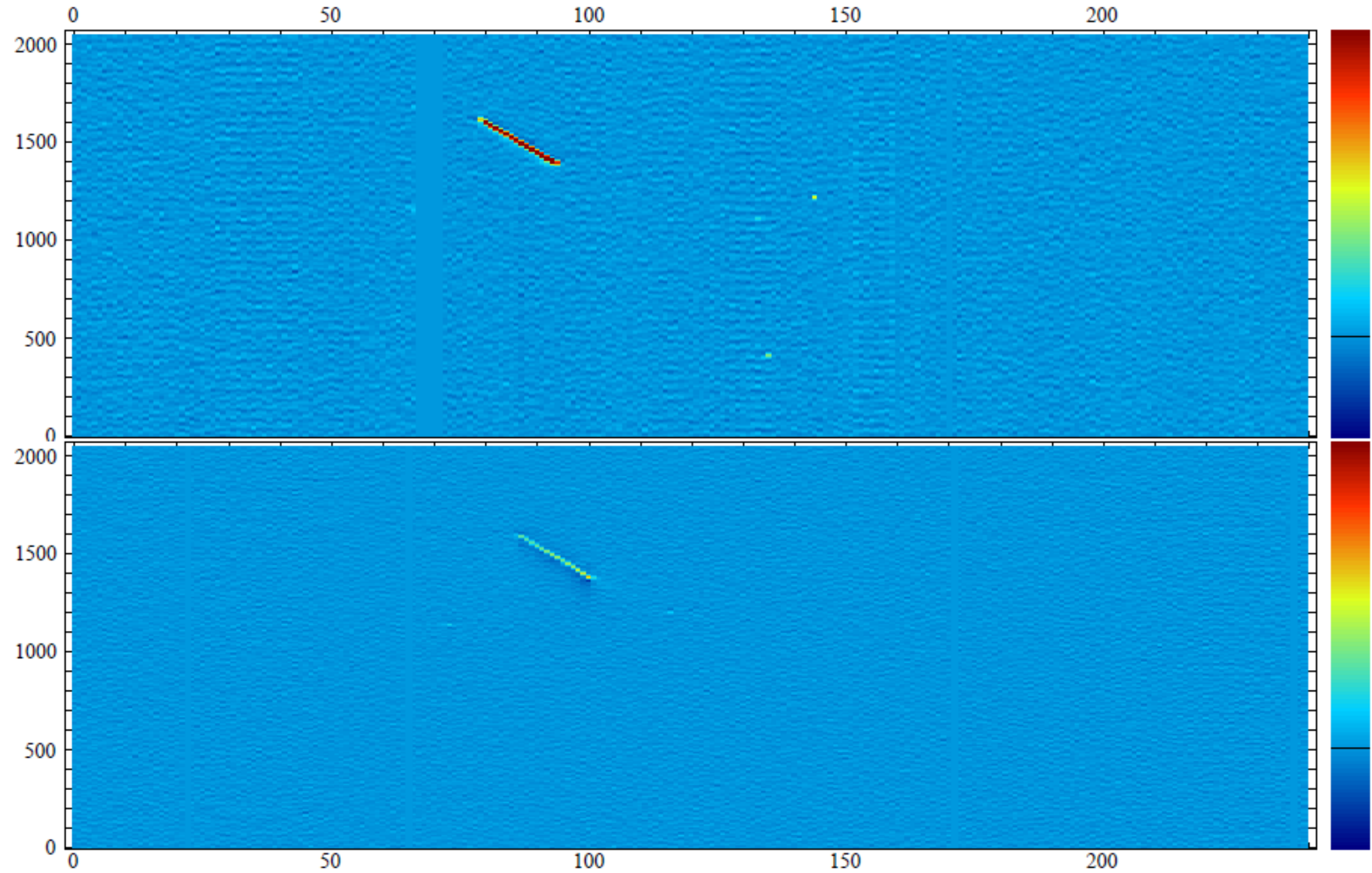}
\caption{LArSoft event display of a fully contained proton with kinetic energy of 100 MeV (data). The wire number of the collection plane (top) and induction plane (bottom) is shown on the horizontal axes. The ArgoNeuT wire spacing is 4 mm. The drift time (0.2 $\mu$s increments) is shown on the vertical axes. The track has 16 space points and a range of 9.7 cm. The track $\phi$ angle is 43$^\circ$.}
\label{fig:track}
\end{figure}

\begin{figure} [H]
\centering
\includegraphics [width = 0.45\textwidth] {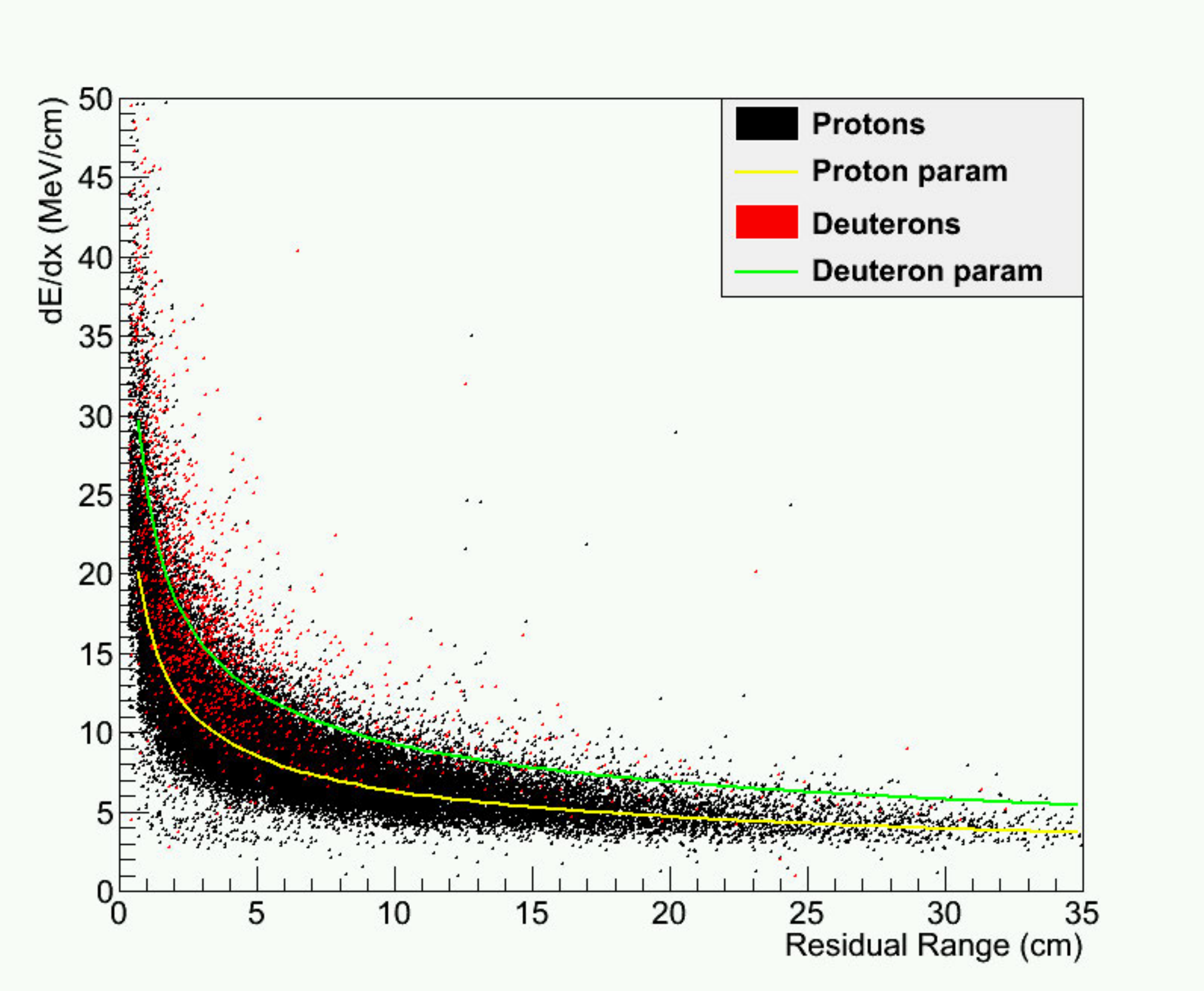}
\includegraphics [width = 0.45\textwidth] {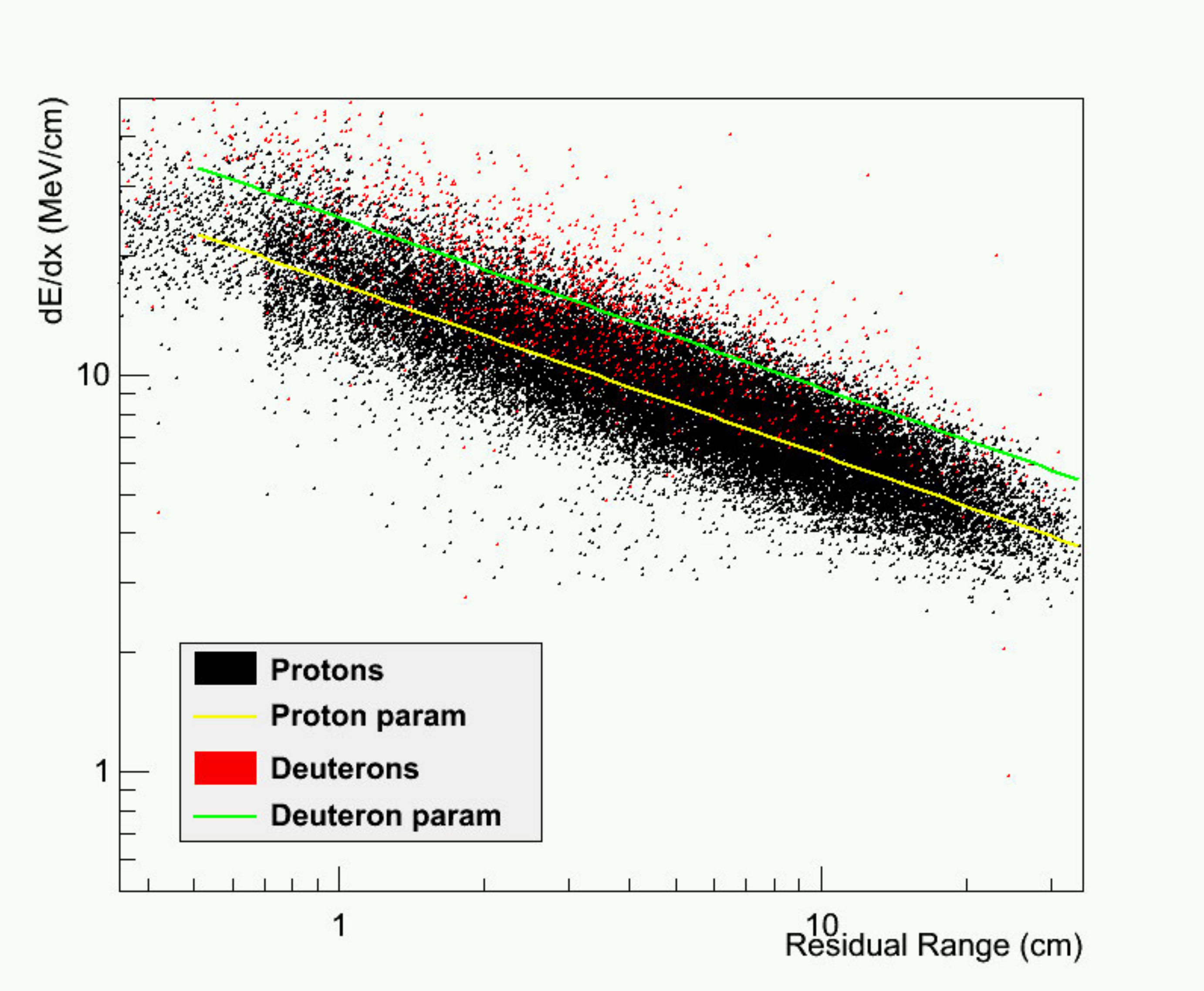}
\caption{Left: Histogram of $(dE/dx)_{calo}$ vs residual range for the proton sample (black) and the deuteron sample (red). Equation \protect\ref{dedxpar} is plotted for protons (yellow line) and deuterons (green line). Right: The same histogram plotted on a log-log scale.}
\label{fig:dedx_resrange}
\end{figure}

We estimate the background in the proton sample by fitting the PIDA distribution to three Gaussian distributions. The fitted values of the mean positions are close to the expected values for pions, protons and deuterons. The pion and deuteron Gaussian distributions are extrapolated into the proton selection region to estimate the proton selection purity ($\approx$95\%) and efficiency ($\approx$95\%). Proton candidate tracks range in length from 3 cm to 30 cm corresponding to a kinetic energy range of 50 MeV to 250 MeV. A typical track is shown in figure \ref{fig:track}.
Figure \ref{fig:dedx_resrange} illustrates the power law behavior for the two samples.

\section{Detector Simulation}

The LArSoft simulation is based on GEANT4. The TPC is divided into 0.3~mm cubic voxels imposing a maximum tracking step size $\approx$10x smaller than the wire spacing. At each tracking step, $dE/dx$ is calculated using the step information and a Birks recombination correction applied to find $dQ$ using equation \ref{birks}. The recombination correction does not include any angular dependence. The deposited charge is split into clusters of 600 electrons which are each subjected to a simulation of longitudinal diffusion, transverse diffusion and loss due to impurities. The number of arriving electrons is stored in a time ordered array for each wire. When tracking of all particles in the event is completed, wire signal waveforms are generated by convolving the wire time arrays with a parameterized induction (collection) plane response and the electronics impulse response.

\begin{figure} [H]
\centering
\includegraphics [width = 0.5\textwidth] {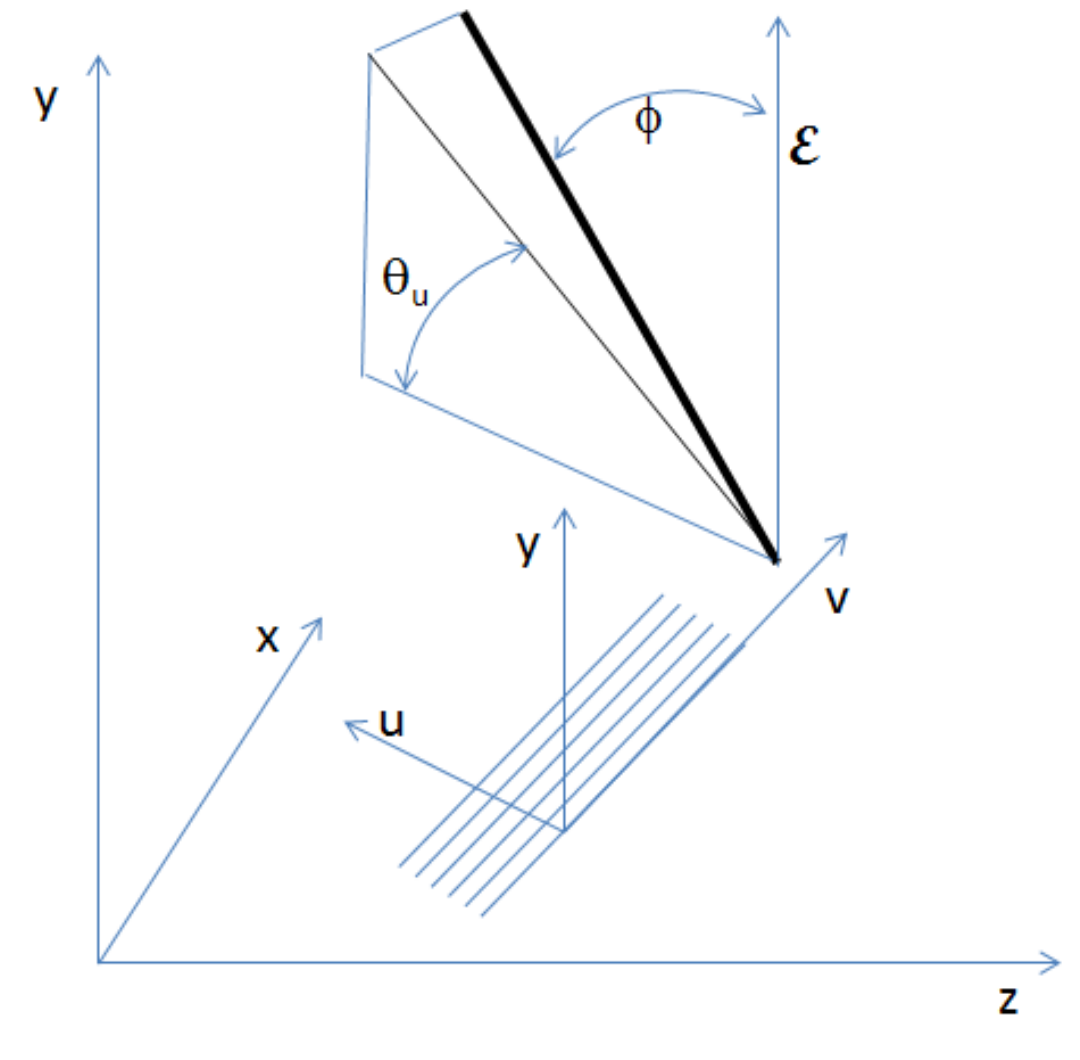}
\caption{Projection of a track (black line) into the (u,y,v) coordinate system of the collection plane wires. The electric field is in the y direction and is perpendicular to the wire plane. The wire plane lies in the (x,z) plane.}
\label{fig:coordsys}
\end{figure}

The Monte Carlo simulation is used to check the relative calibration of tracks at all $\phi$ angles. The rationale for such an effect becomes clear by considering two minimum ionizing tracks, both traveling at the same angle relative to the collection plane wires. One track travels parallel to the wire plane ($\phi$ = 90$^\circ$) and the other is inclined relative to the wire plane. The case of an inclined track is shown in figure \ref{fig:coordsys}, where the inclination angle $\theta_u$ is the angle between the projection of the track in the (u,y) plane and the u axis. The wire plane lies in the (x,z) plane. The inclination angle is highly anti-correlated with $\phi$, e.g. $\phi$ = 90$^\circ$ when $\theta_u$ = 0 in the parallel track case. Ionization electrons arrive at the collection plane with a larger spread in time for the inclined track case than for the parallel track case. A hit signal from the inclined track will therefore have a wider time spread and lower amplitude but the total charge deposited per unit length, $dQ/dx$, will be the same for both tracks. The power spectrum of the wire signal from the inclined track has lower frequency components than the wire signal from the parallel track. The signals for the two cases may therefore be processed differently by the readout electronics chain, signal deconvolution and hit fitting, resulting in a slightly different measurement of  $dQ/dx$.

We investigate the need for an angle dependent calibration using the Monte Carlo. We simulate minimum ionizing muons with a fixed momentum at a wide range of angles and pass the simulated wire signals through the model of the electronics chain. Hits and tracks are then reconstructed using the same calorimetry code that is used to analyze the data. This study shows the need to apply a calibration correction factor (1 + 0.048 $ \times \cos^2(\theta_{\rm u})$) to $dQ/dx$. This is a 3\% correction at $\theta_{\rm u}$ = 40$^\circ$. We confirm that the measurement of $dQ/dx$ is independent of $\theta_{\rm u}$ after the correction term is applied.

We next confirm that this calibration correction does not introduce an artificial angular dependence in the recombination analysis. Simulated stopping protons are generated with similar distributions in angle and kinetic energy as the data. The simulated data are subjected to the same hit reconstruction, track reconstruction and calorimetric analysis as the real data. We find that the ratio of $(dQ/dx)_{\phi = 40^\circ} / (dQ/dx)_{\phi = 80^\circ}$ vs $(dE/dx)_{hyp}$  is 1.00 $\pm$ 0.02. This result is expected since the recombination model used in the Monte Carlo does not include any angular dependence. It demonstrates that the angle dependent calibration is correctly applied. We conclude that any angular dependence observed in the data is not an artifact of reconstruction, calibration or the analysis method.

\section{Recombination Simulation}

Jaskolski and Wojcik provided a theoretical-computational approach to the columnar model in \cite{jaskolski} to test the validity of the theoretical premises. In this simulation, positive ions are uniformly distributed along a line representing the track trajectory. The ions are separated by a distance $r_{ion}$ and are considered to be immobile. An electron is generated randomly about each ion with defined spatial and kinetic energy distributions. The electrons are then allowed to move in time steps under the influence of the combined Coulomb field of all ion-electron pairs and an external electric field that is perpendicular to the track trajectory ($\phi$ = 90$^\circ$). The simulation results are insensitive to the assumptions used for the initial electron distributions because the magnitude of the initial state conditions is small compared to conditions after thermalization. Interactions with the surrounding argon atoms are modeled by energy-transfer and momentum-transfer cross sections. Electron-ion pairs are removed from the simulation, i.e. they recombine, if their separation distance is less than 1.3 nm. Electrons escape recombination if they drift more than  a distance $y_{max}$ = 4000 nm from the track. The results of this simulation are in quantitative agreement with the ICARUS data at high $dE/dx$. The poorer agreement at low $dE/dx$ can be rectified by introducing a factor that accounts for energy loss due to $\delta$-rays which are not included in the model. 

\begin{figure} [H]
\centering
\includegraphics [width = 0.5\textwidth] {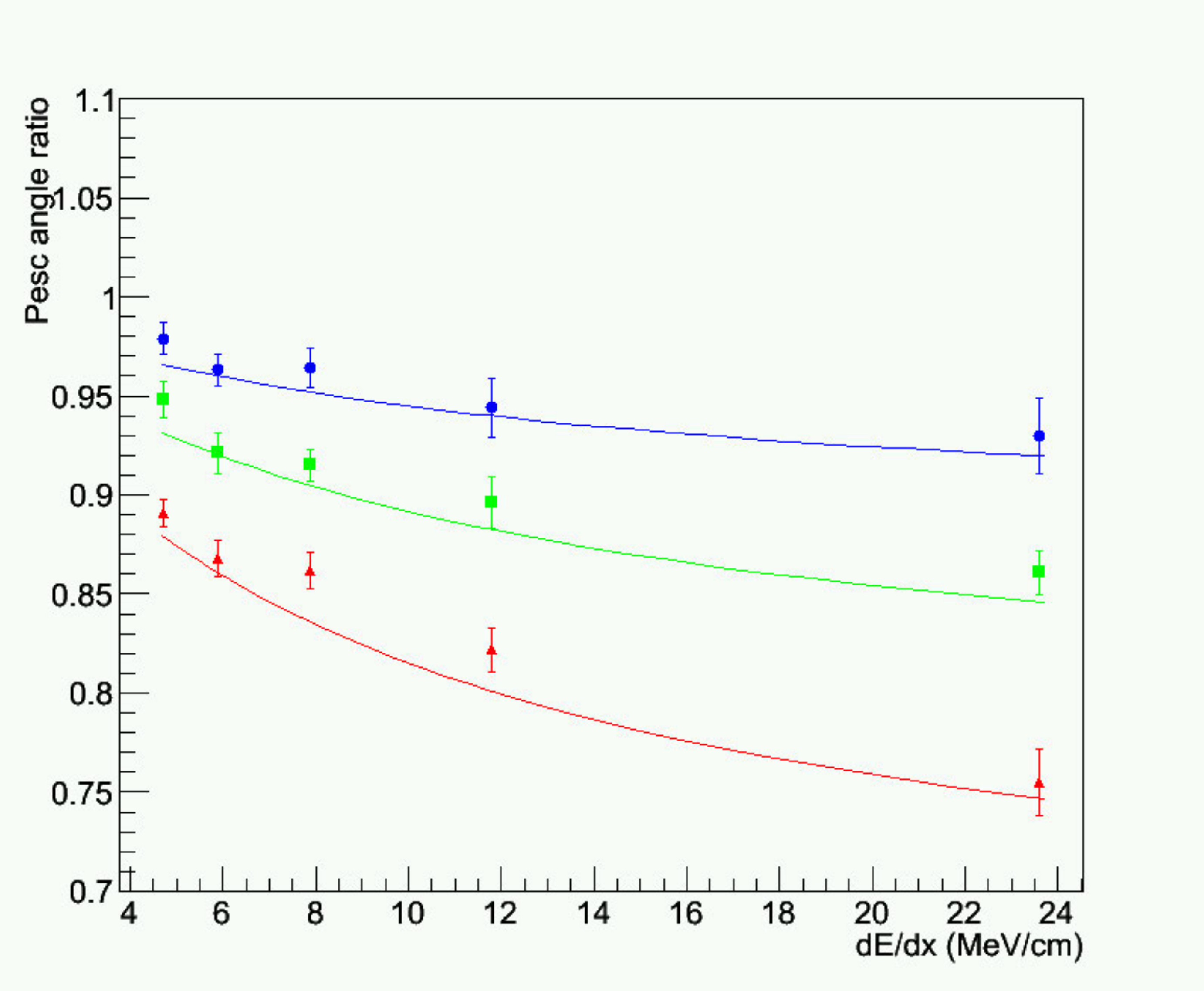}
\caption{Ratios of escape probabilities from the recombination simulation (points) for $\mathcal{R}_{60^\circ}$ / $\mathcal{R}_{80^\circ}$ (blue),  $\mathcal{R}_{50^\circ}$ / $\mathcal{R}_{80^\circ}$ (green) and $\mathcal{R}_{40^\circ}$ / $\mathcal{R}_{80^\circ}$ (red). The colored lines represent the angular dependence of the columnar theory. }
\label{fig:recombsim}
\end{figure}

This work has been extended by Wojcik to simulate recombination from tracks with $\phi$ < 90$^\circ$. Studies show that the recombination factor becomes independent of $y_{max}$ for tracks at $\phi$ = 40$^\circ$ when $y_{max} \geq$  7000 nm. In this work, we set $y_{max}$ = 7000 nm for all angles and run the simulation for 10000 electrons. The simulation is performed for values of $r_{ion}$ = 10, 20, 30, 40 and 50 nm and $\phi$ = 40$^\circ$, 50$^\circ$, 60$^\circ$ and 80$^\circ$. The results are shown in figure \ref{fig:recombsim} as ratios of  $\mathcal{R}_{\phi} / \mathcal{R}_{80^\circ}$ as a function of $dE/dx$ (= $W_{ion}/r_{ion}$). The prediction of the columnar theory is shown by the curved lines where we have replaced $\mathcal{E}$ with $\mathcal{E}$sin$\phi$ in equation \ref{birks}. The angular dependence from theory and simulation is both consistent and significant, for example with a charge loss of 25\% at $dE/dx$ = 24 MeV/cm and $\phi$ = 40$^\circ$ relative to a stopping proton traveling perpendicular to the electric field.

\section{Analysis and Results}

\subsection{Proton Sample}

\begin{figure} [H]
\centering
\includegraphics [width = 0.7\textwidth] {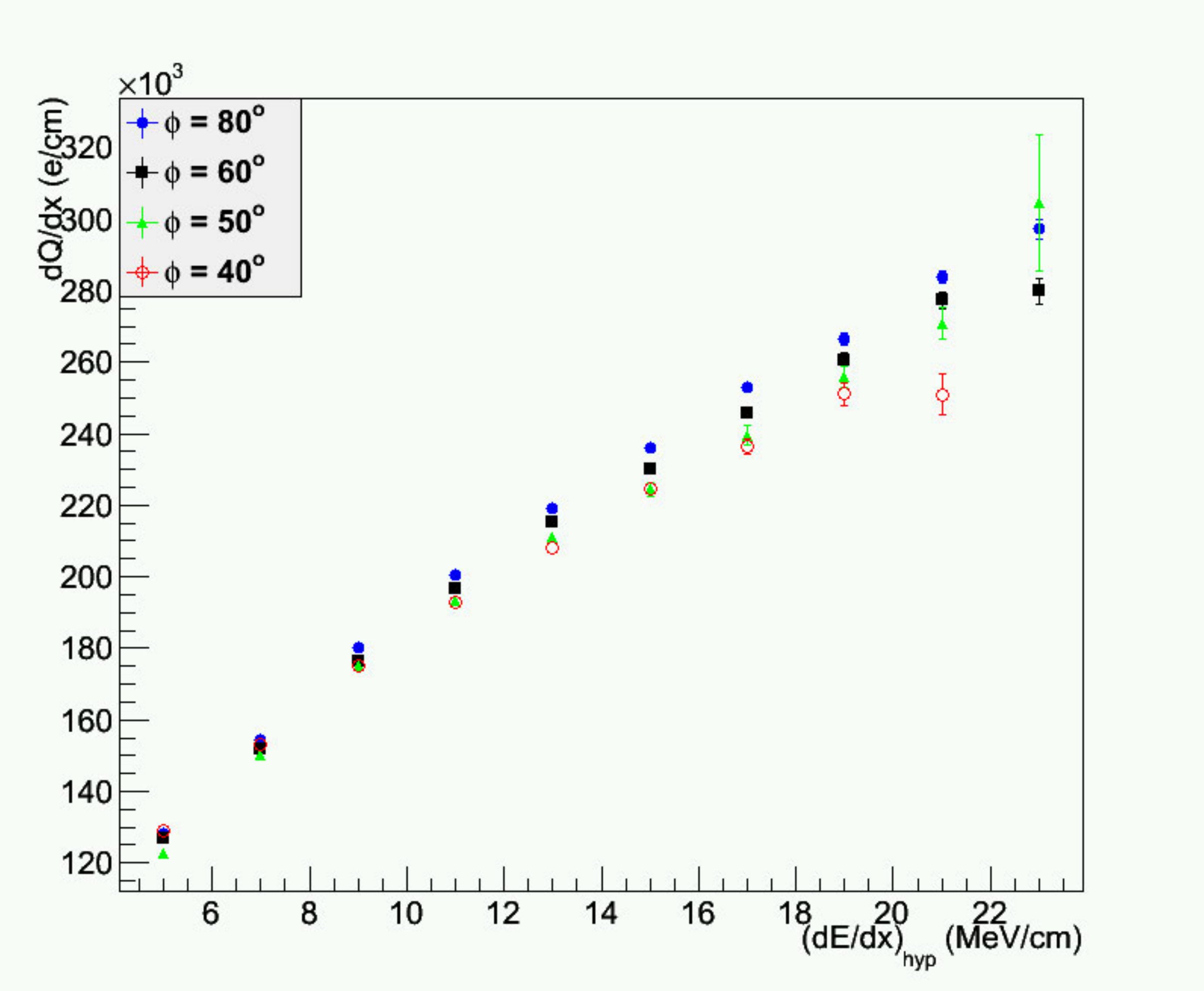}
\includegraphics [width = 0.7\textwidth] {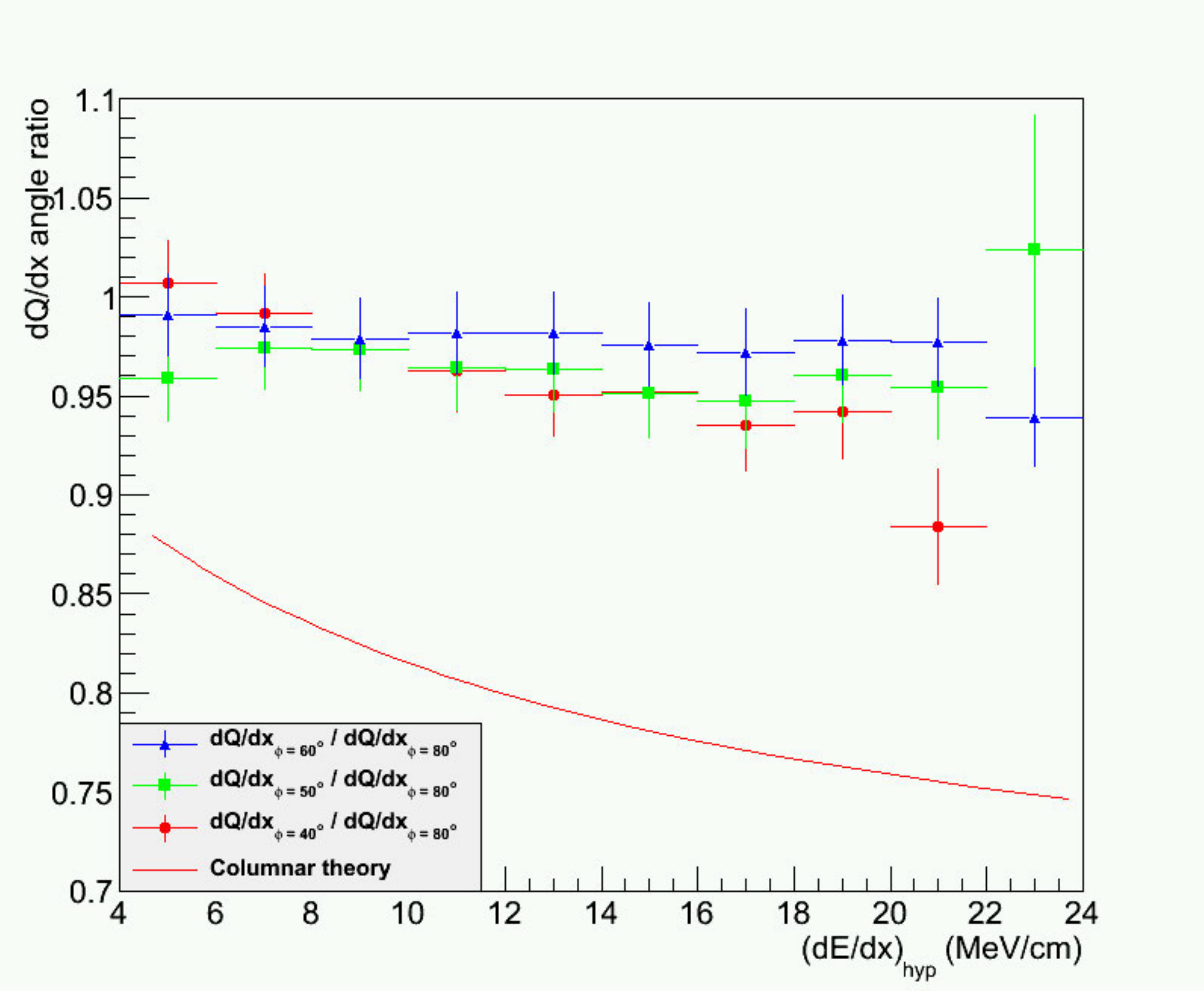}
\caption{Top: $dQ/dx$ vs $(dE/dx)_{hyp}$ for all angle bins. The vertical bars represent the statistical error on $dQ/dx$. Bottom: Ratios of the data in the top plot (data points) including a 2\% systematic error on $dQ/dx$. The red curve is the expectation of the columnar theory, reproduced from figure \protect\ref{fig:recombsim}.}
\label{fig:dqdx_all}
\end{figure}

The Birks and Box recombination parameters are found by fitting histograms of $dQ/dx$ vs $(dE/dx)_{hyp}$. These variables are chosen since neither is subject to any recombination model assumptions. The values for these variables are accumulated in 2 MeV/cm $(dE/dx)_{hyp}$ bins and 4 $\phi$ bins for space points that are within the fiducial volume. It is apparent by inspection of the data in figure \ref{fig:dqdx_all} that the dependence on $\phi$ is significantly weaker than the theory predicts. The difference between the data points in the lowest $(dE/dx)_{hyp}$ bin is a few \%. Recombination reduces the collected charge by 5\% - 10\% at small $\phi$ and high ionization. This is significantly less than the 25\% loss predicted by the Jaffe columnar theory and the recombination simulation.

\begin{figure} [h!]
\centering
\includegraphics [width = 0.48\textwidth] {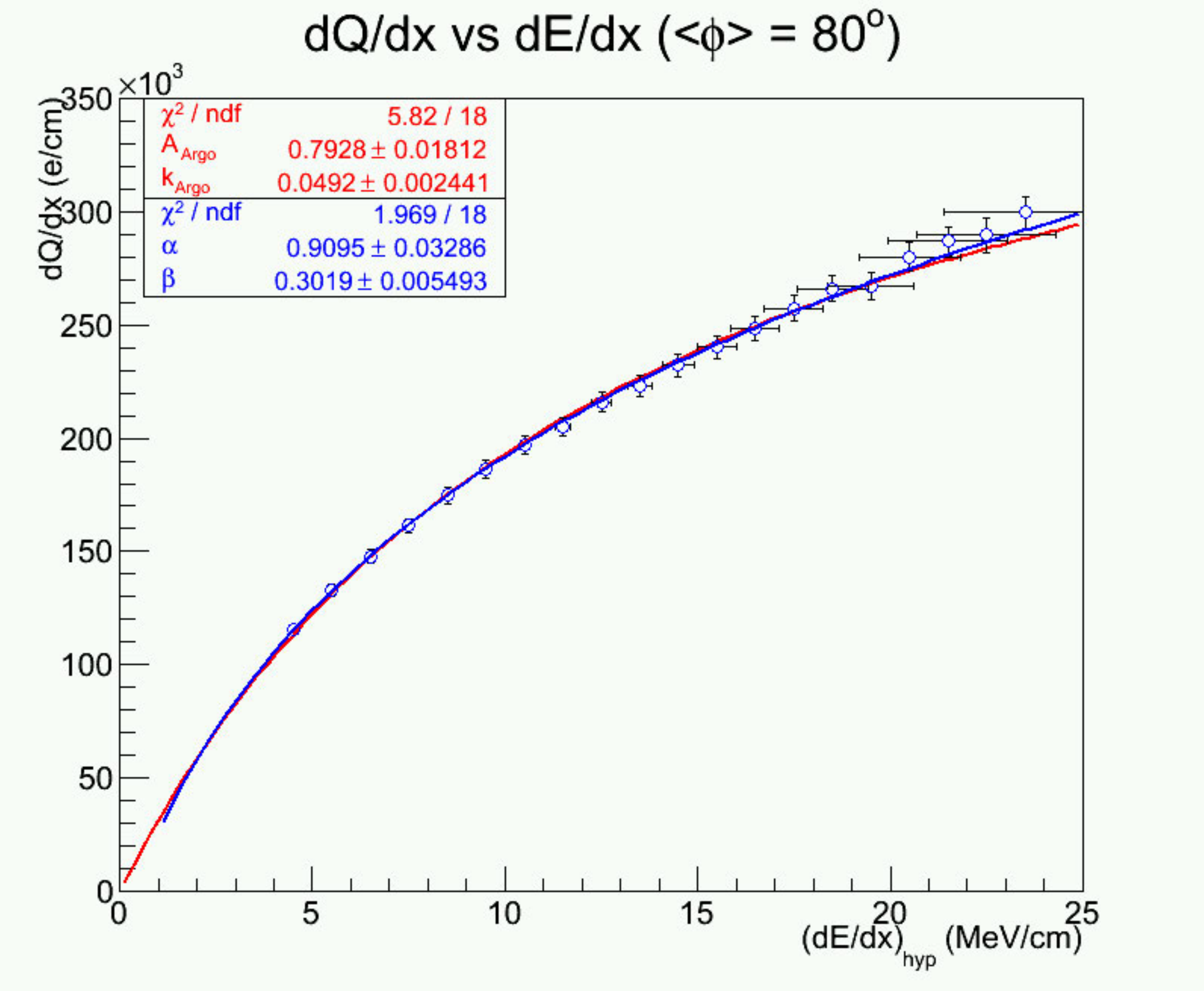}
\includegraphics [width = 0.48\textwidth] {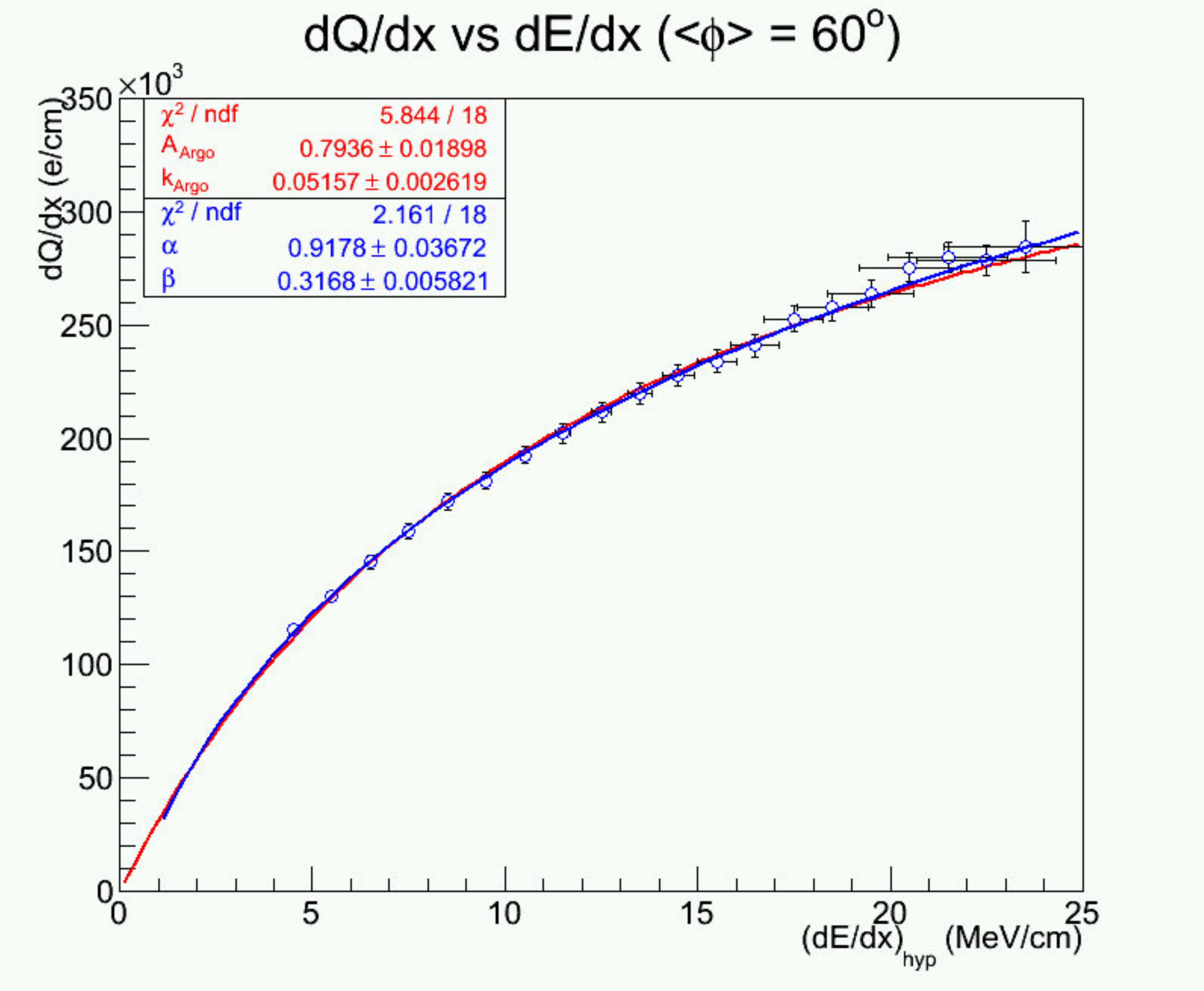} \\
\includegraphics [width = 0.48\textwidth] {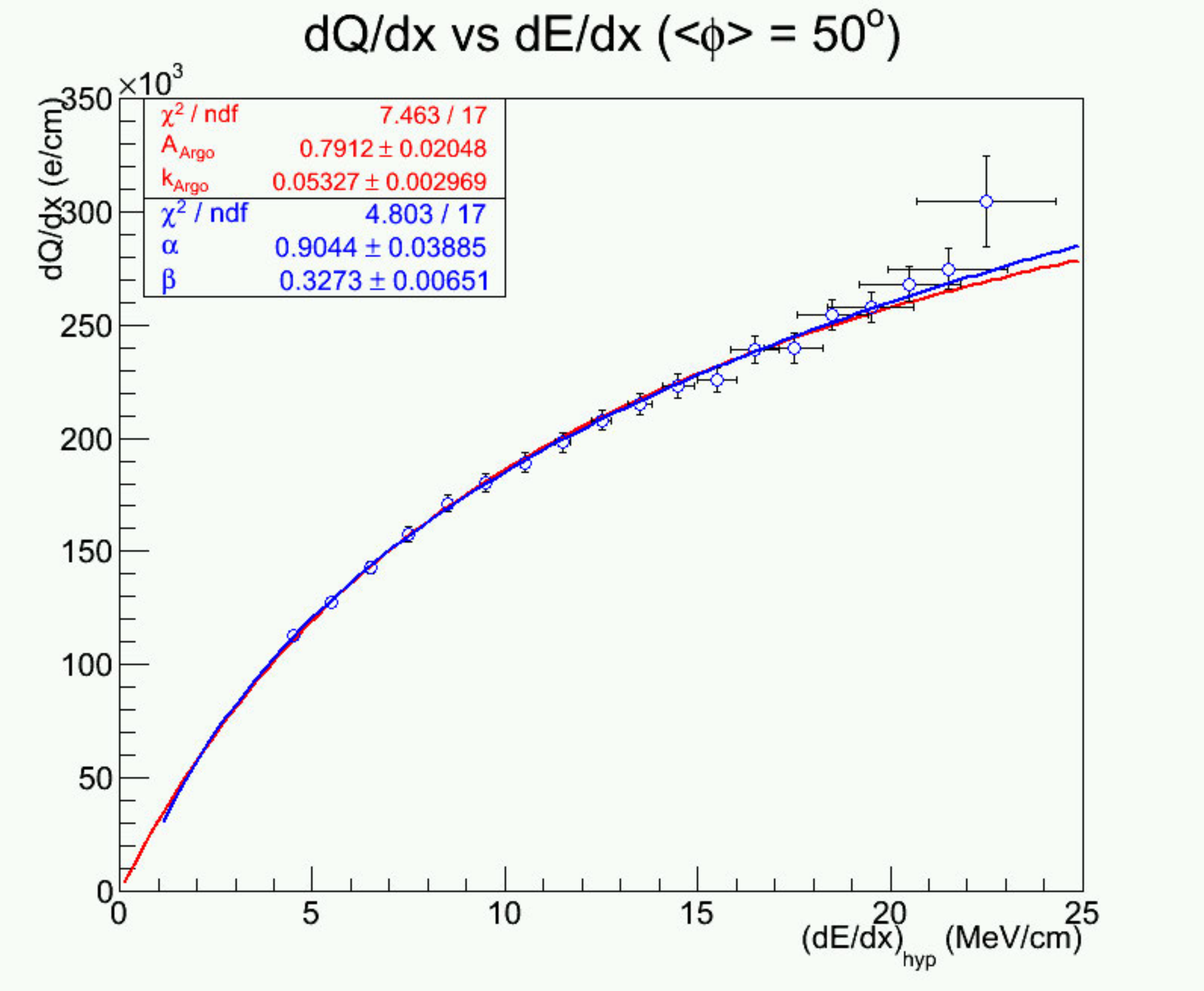}
\includegraphics [width = 0.48\textwidth] {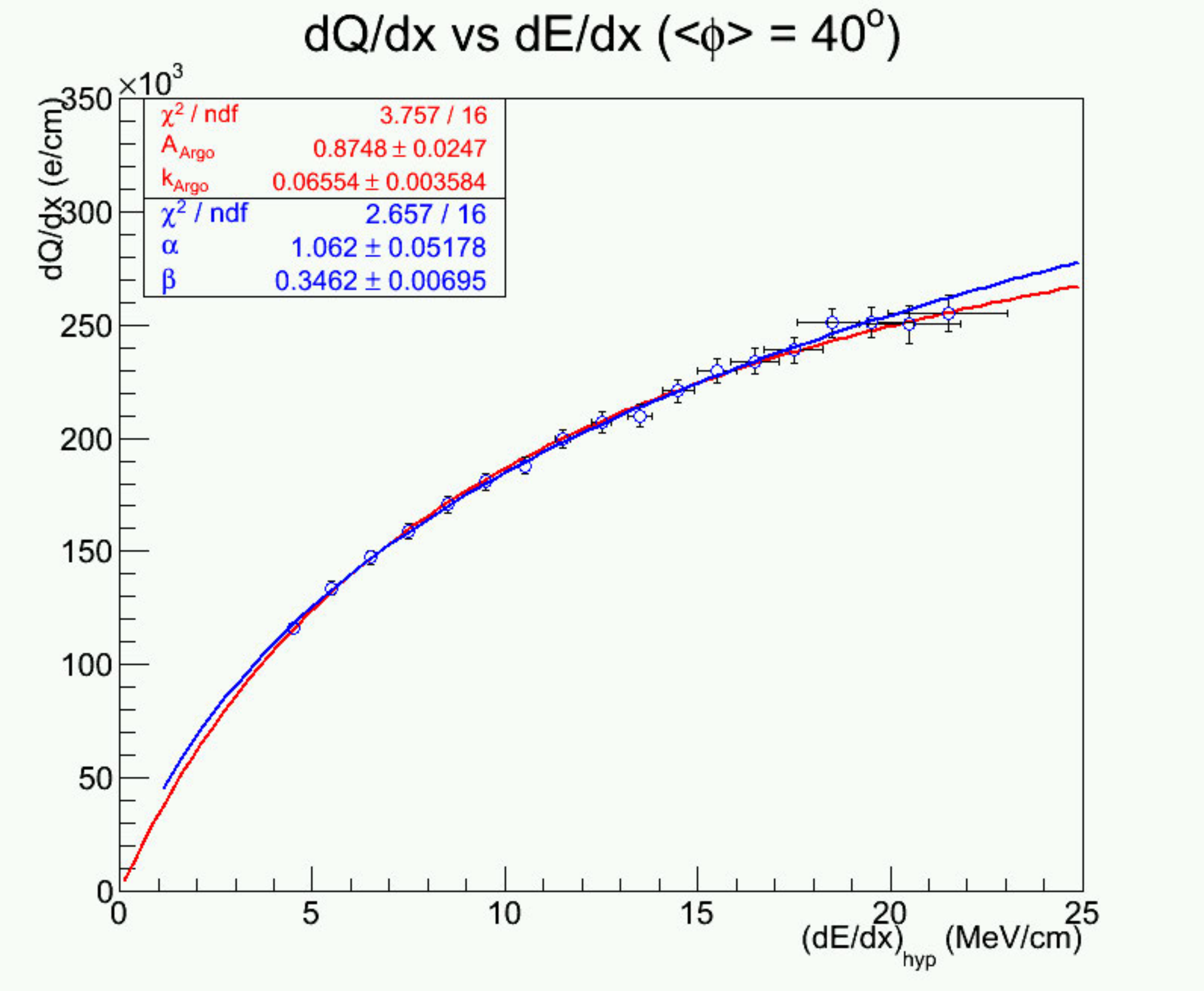}
\caption{Recombination fits to $dQ/dx$ vs $(dE/dx)_{hyp}$ for all angle bins. The red (blue) curves and text represent Birks model (modified Box model) fits.}
\label{fig:allfit}
\end{figure}

Prior to fitting, a 2\% systematic error is added in quadrature with the statistical errors. The stopping point error results in an uncertainty on the residual range, $R$, for each space point and therefore on $(dE/dx)_{hyp}$. The error on $(dE/dx)_{hyp}$ is found by propagating the 0.1 cm residual range error (Section \ref{sec:caloreco}) using equation \ref{dedxpar}. The recombination fits in each angle bin are shown in figure \ref{fig:allfit}. The results are summarized in Table \ref{fitsummary}. 

\begin{table} [H]
\centering
\begin{tabular} {| c | c | c | c | c | c |}
\hline
Angle Bin & Angle Bin  & Box $\alpha$  & Box $\beta$             & Birks A$_{\rm{Argo}}$ & Birks k$_{\rm{Argo}}$ \\ 
      &    Range      &                      &  (MeV/cm)$^{-1}$  &           &  (kV/cm)(g/cm$^2$)/MeV \\
\hline \hline
80$^\circ$ & 70$^\circ$ - 90$^\circ$  & 0.91 $\pm$ 0.03  & 0.302 $\pm$ 0.005 & 0.793 $\pm$ 0.018 & 0.049 $\pm$ 0.002 \\
60$^\circ$ & 55$^\circ$ - 70$^\circ$  & 0.92 $\pm$ 0.04  & 0.317 $\pm$ 0.006 & 0.794 $\pm$ 0.019 & 0.052 $\pm$ 0.003 \\
50$^\circ$ & 47$^\circ$ - 55$^\circ$  & 0.90 $\pm$ 0.04  & 0.327 $\pm$ 0.007 & 0.791 $\pm$ 0.020 & 0.053 $\pm$ 0.003 \\
40$^\circ $& 20$^\circ$ - 47$^\circ$  & 1.06 $\pm$ 0.05  & 0.346 $\pm$ 0.007 & 0.875 $\pm$ 0.025 & 0.066 $\pm$ 0.004 \\
\hline
 All            & 20$^\circ$ - 90$^\circ$  & 0.93 $\pm$ 0.02  & 0.319 $\pm$ 0.003 & 0.806 $\pm$ 0.010 & 0.052 $\pm$ 0.001 \\\hline 
\end{tabular}
\caption{Summary of Birks and modified Box model fits for the proton sample.}
\label{fitsummary}
\end{table}

The values of the Birks parameters A$_{\rm Argo}$ and k$_{\rm Argo}$ found in this analysis are in excellent agreement with the ICARUS values at $\phi$ = 80$^\circ$, $A_B$ = 0.800 $\pm$ 0.003 and $k_B$ = 0.0486 $\pm$ 0.0006 (kV/cm) (g/cm$^2$)/MeV. Both the Birks and modified Box model equations provide a good representation of the data in the range 2 MeV/cm < $dE/dx$ < 24 MeV/cm. 

A modified form for $\xi$ that removes the dependence on the ArgoNeuT operating conditions is

\begin{equation}
\label{betaprime}
\xi = \beta' (dE/dx) / (\rho \mathcal{E}) 
\end{equation}

\noindent
The value of $\beta'$ for the angle range 20$^\circ$ - 90$^\circ$ is $\beta' =  0.212 \pm 0.002 \; \rm{(kV/cm)(g/cm^2)/MeV}$. The expectation that the value of $\alpha$ in the modified Box model is independent of angle is met except for the 40$^\circ$ bin. The weighted average value is $\alpha$ = 0.93 $\pm$ 0.02 for all angle bins. 

For comparison, an extrapolation of the modified Box model and Birks model curves to the stopping power for a minimum ionizing particle results in a negligible 0.2\% difference. Therefore, although both models fit the data well, the inverse modified Box model used to calculate $(dE/dx)_{calo}$ behaves better at high ionization density.

\subsection{Deuteron Sample}

\begin{figure} [H]
\centering
\includegraphics [width = 0.48\textwidth] {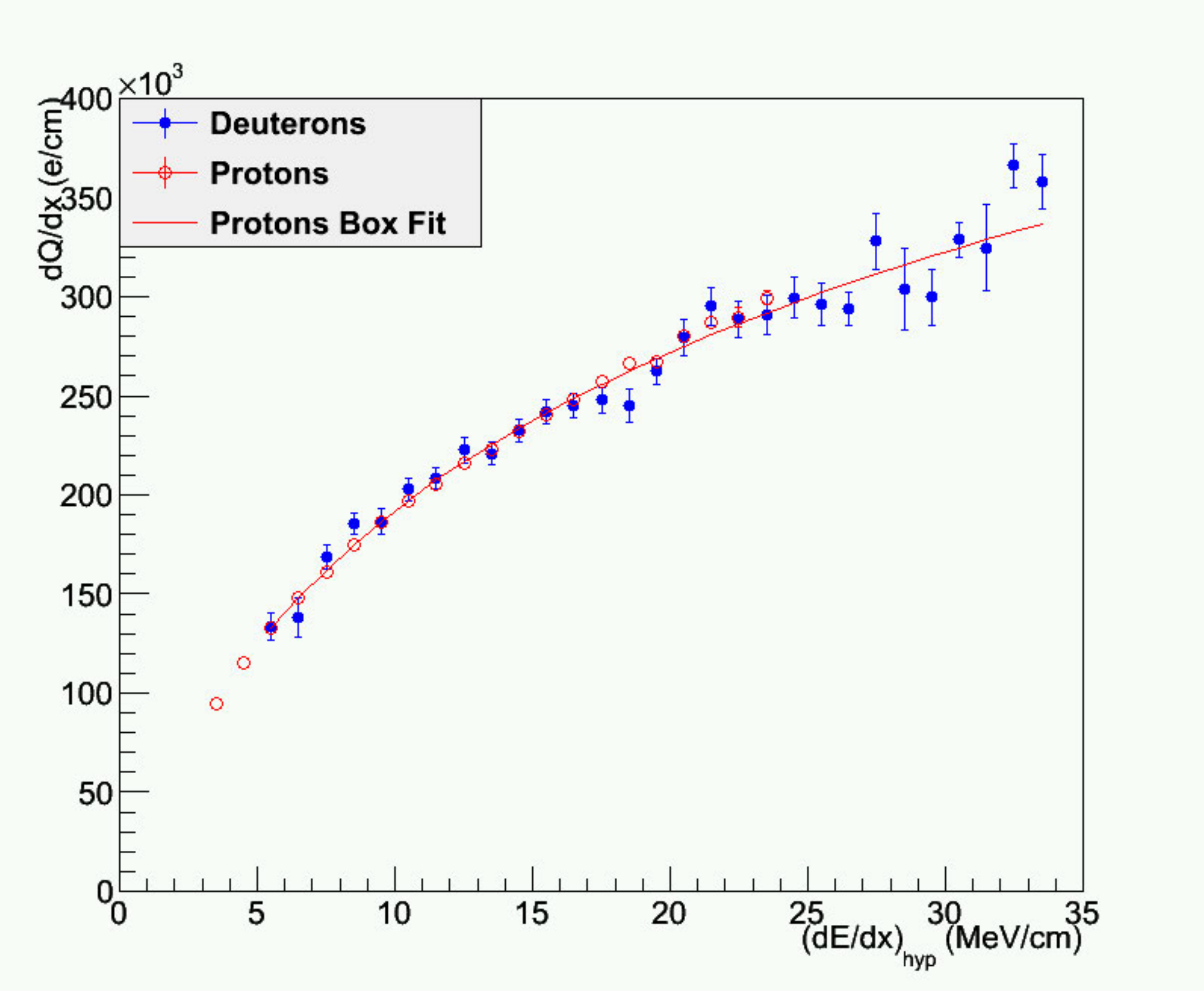}
\includegraphics [width = 0.48\textwidth] {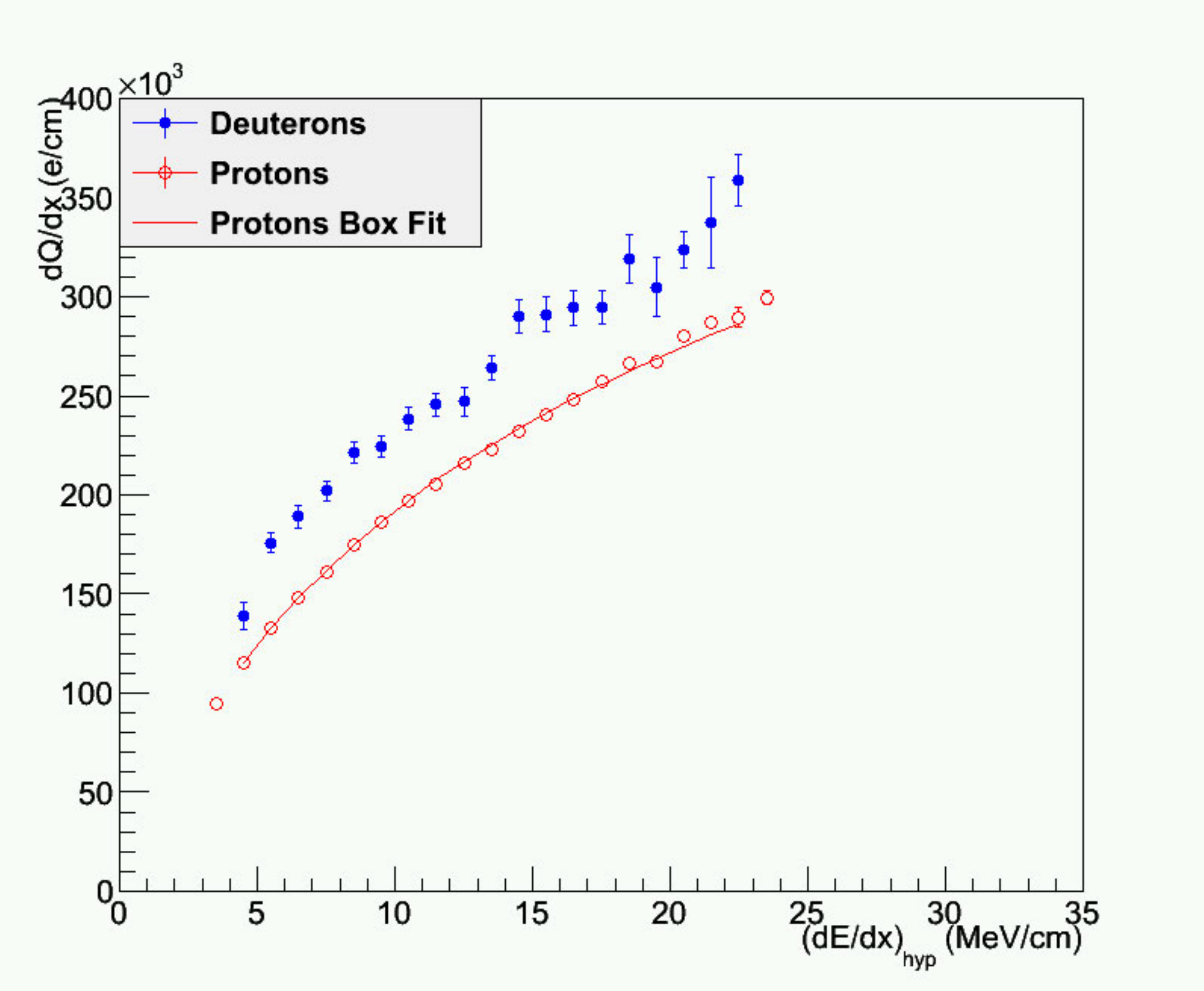}
\caption{Left: Recombination in the deuteron sample compared with protons for $\phi$ =80$^\circ$ using a deuteron $(dE/dx)_{hyp}$ hypothesis. Right: Same as the left plot but using a proton hypothesis for the deuteron sample.}
\label{fig:deutrecomb}
\end{figure}

Recombination at higher stopping power can in principle be studied with deuterons. The small number of tracks in our sample limits its usefulness however. The deuteron sample is subjected to the stopping point fitting algorithm described above with $(dE/dx)_{hyp}$ calculated using the deuteron hypothesis. The deuteron data for the $\phi$ = 80$^\circ$ bin are shown in blue in figure  \ref{fig:deutrecomb}. The red points and curve are the data and modified Box fit parameterization for protons in the same $\phi$ bin. The $\chi^2$ agreement between the deuteron data and the modified Box fit is 31.6 with 29 degrees of freedom in the left hand figure. The deuteron data are incompatible with a proton hypothesis as the right-hand figure shows. We conclude that deuterons are indeed present in this sample and that the recombination fits found above are applicable to $dE/dx$ = 35 MeV/cm.

\section{Discussion}

As noted earlier, a theoretical understanding of recombination should result in a consistent prescription for calorimetric reconstruction. The significant disagreement between published data and theory belies that statement however. One recognized difference between theory and practice is the presence of $\delta$-rays. We propose that another significant difference is the presence of impurities in a real detector. Studies at the Fermilab Materials Test System have shown that the primary contaminant is water \cite{mts}. We  consider the possibility that charges attached to water molecules could create an electrostatic screening effect that would mitigate the angle dependence. Electrostatic screening plays a role in plasma physics and in electrolytes and is characterized by the ``Debye length''.  A calculation of the Debye length using the measured drift electron lifetimes in the ArgoNeuT detector results in a screening length of 400 - 600 nm. This is four times larger than the Onsager length \cite{onsager} and should therefore not be a significant contribution. We have looked for such an effect in the data by comparing the ratio $(dQ/dx) / (dE/dx)_{hyp}$ for $(dE/dx)_{hyp}$ > 18 MeV/cm in all angle bins as a function of the electron lifetime and find that the ratios do not depend on the argon purity.

We find that the angular dependence of the electron escape probability becomes weaker as the parameter $y_{max}$ is decreased in the recombination simulation. For example, at $dE/dx$ = 24 MeV/cm with $y_{max}$ = 2000 nm, the escape probability at $\phi$ = 40$^\circ$ is only 13\% smaller than the value at $\phi$ = 80$^\circ$, and is in reasonable agreement with our data. This result can be understood by noting that the electron thermalization distance in liquid argon is $\sim$2500 nm\cite{sowada}, so the electrons that are assumed to escape recombination at $y_{max}$ = 2000 nm have rather high kinetic energies. Therefore, their motion should not be strongly affected by the direction of an applied external field. On the other hand, the assumption that electrons become effectively independent of their original tracks when they separate from them by only $\sim$2000 nm is not  unrealistic in a real detector environment. This may be due to the presence of positive and negative charges that are left from different ionization events. The above reasoning provides a possible explanation of the weak angular dependence of the recombination factor observed in the ArgoNeuT experiment. More theoretical studies are needed to fully understand the electron recombination processes in a real liquid argon detector.

\section{Conclusions}

We have studied the angular dependence of electron recombination in a large sample of stopping protons in the ArgoNeuT detector.  The angular dependence is significantly weaker than that predicted by the Jaffe columnar theory and by a recombination simulation. As noted above, this is not the sole example of disagreement with recombination models. We have presented two possibilities to explain this discrepancy. Both possibilities have a common theme - that impurities in the argon may affect the micro-physics of electron recombination. 

\acknowledgments

We gratefully acknowledge the support of Fermilab, the U.S. Department of Energy and the National Science foundation. Fermilab is operated by Fermi Research Alliance, LLC under Contract No. DE-AC02-07CH11359 with the United States Department of Energy.

\end{document}